\documentclass[twocolumn,showpacs,aps,prb,superscriptaddress]{revtex4}

\usepackage{graphicx}
\usepackage{dcolumn}
\usepackage{epsfig}
\usepackage{bm}

\begin{document}

\preprint{}

\title{First-principles study of structural, elastic, and bonding properties of pyrochlores}

\author{J. M. Pruneda}
\affiliation{Department of Earth Sciences, University of Cambridge
Downing Street, Cambridge, CB2 3EQ, United Kingdom}
\affiliation{Institut de Ci\`encia de Materiales de Barcelona (CSIC), Spain.}
\author{Emilio Artacho}
\affiliation{Department of Earth Sciences, University of Cambridge
Downing Street, Cambridge, CB2 3EQ, United Kingdom}
\affiliation{Donostia International Physics Center (DIPC),
Paseo Manuel de Lardizabal 4, 20018 Donostia-San Sebasti\'an, Spain}

\date{\today}

\begin{abstract}
Density Functional Theory calculations have been performed to obtain 
lattice parameters, elastic constants, and electronic properties 
of ideal pyrochlores with the composition A$_2$B$_2$O$_7$ (where A=La,Y 
and B=Ti,Sn,Hf, Zr).  Some thermal properties are also inferred from the 
elastic properties.  A decrease of the sound velocity (and thus, of the 
Debye temperature) with the atomic mass of the B ion is observed.  Static and
dynamical atomic charges are obtained to quantify the degree of 
covalency/ionicity.  A large anomalous contribution to the dynamical 
charge is observed for Hf, Zr, and specially for Ti.  It is attributed 
to the hybridization between occupied $2p$ states of oxygen and 
unoccupied d states of the B cation.  The analysis based on Mulliken 
population and deformation charge integrated in the Voronoi polyhedra
indicates that the ionicity of these pyrochlores increases in the 
order Sn--Ti--Hf--Zr.  The charge deformation contour plots support 
this assignment.
\end{abstract}


\maketitle

\section{Introduction}

Materials with the pyrochlore (A$_2$B$_2$O$_7$) lattice structure have 
unique properties that make them ideal candidates for applications 
ranging from high-permittivity dielectrics\cite{dielectrics}, 
to ceramic thermal barrier coatings (TBC)\cite{thermal},
potential solid electrolytes in solid-oxide fuel cells\cite{fuelcell}, 
or immobilization hosts of actinides in nuclear waste\cite{waste}.
Many studies have tried to optimize the pyrochlore composition to obtain 
the most desirable properties for a particular application.  
One has to achieve maximum efficiency in a special
property (high dielectric constant, low thermal conductivity, 
low migration energies, etc.), but at the same time satisfy other 
requirements such as physical stability, chemical or thermal compatibility 
with other materials involved in the design, etc.  In many cases, no 
experimental information is available, or if there is information there is no
agreement between different experiments, a consequence of the 
difficulties in growing pure samples.

Therefore, simulations have become an ideal 
tool for systematic investigations of various properties as a function of the 
chemical composition\cite{Sickafus00,TBC,Tabira2000,migration,Chartier02}. 
Most of the atomistic simulations studies are based in classical 
potential methods, where the interatomic interactions are parametrized 
with effective potentials fitted to reproduce experimentally 
known data, as the unit cell dimensions of a variety of oxide pyrochlores.
The complexity of the chemical elements involved (mostly transition 
metals and rare earths) makes desirable the use of {\it ab initio} 
calculations, but the large size of the unit cell in pyrochlores (88 atoms) 
had restricted the application of this technique.  

The limitations of these classical models have recently been pointed out 
by Panero et. al.\cite{Panero04} in the study of the energetics of 
cation-antisite defects.  The stability of pyrochlores and the resistance 
to amorphization by irradiation has been correlated with the propensity 
of the ions to create these point defects\cite{Sickafus00}.  
With classical models, the defect-formation energies show a dependence 
on the cationic radius: compounds with similar radii for A and B are 
more radiation resistant than compounds with very dissimilar radii, 
that show greater susceptibility to amorphization.  The effect of the 
B cation radius seems to be more important than the effect of the A cation.  
Nevertheless, density functional calculations~\cite{Panero04} have 
shown that the defect formation energies are not simple functions of 
the cationic radius, and a significant influence of the electronic 
configuration of the A and B cations is observed.  

It is now believed that the radiation response cannot be described 
exclusively in terms of the cationic radius ratio and the defect 
formation energies, and that the bond-type must be 
considered.\cite{waste,Lian04,Lian03,Trachenko04,Trachenko}  
It has been suggested that the resistance of materials to amorphization 
of a complex non-metallic compound is determined by a competition between 
short-range covalent interactions, and long-range ionic 
forces.~\cite{Trachenko,Trachenko05}
This picture is based on experimental evidences for more than 100 different 
materials, with very different chemical and structural compositions.  
Unfortunately, it is not easy to quantify the covalency/ionicity and 
relate this to radiation resistance, specially when the
 {\it topological freedom} of the lattice structure can play an 
important role\cite{topological}.  

The broad range of chemical properties 
that can be obtained changing the composition of A- and B-site 
cations in the pyrochlore structure, makes them good testing systems 
to study the relationship between electronic structure and resistance 
to amorphization by radiation damage.  Compounds with strong ionic 
character, like the zirconates (A$_2$Zr$_2$O$_7$) are known to be more 
resistant to radiation to clearly covalent stannates (A$_2$Sn$_2$O$_7$).
Experiments with Gd$_2$(Zr$_x$Ti$_{1-x}$)$_2$O$_7$ show that the 
radiation resistance increases with increasing Zr-content, a clear 
indication of the effect of the electronic structure.

Here we present first principles calculations of two families of 
pyrochlores: La$_2$B$_2$O$_7$ and Y$_2$B$_2$O$_7$, with 
B=Ti, Zr, Sn or Hf with very different bonding properties. The structural
properties and elastic constants are compared with available experimental
information\cite{unstable}, as well as with data obtained with classical atomic 
models.  The structural information obtained with these {\it ab initio} 
calculations can be used to improve the classical interatomic potentials available.

We will discuss the electronic properties of the different compounds and try
to quantify the ionicity/covalency of the different bonds.  Static and 
dynamical charges are frequently used to characterize the nature of bonds
in molecules and solids.  Static charges associated to an isolated atom are
intuitive but ill-defined quantities.  Dynamical charges, on the other side,
are less intuitive, but appear as a more fundamental quantity.  
These dynamical charges are very sensitive to the structural 
properties, but also to partial hybridization between occupied 
and unoccupied states,~\cite{Ghosez98} and hence they can give 
valuable information on the nature of the atomic bonds.

The paper is organized as follows. In Sec.~\ref{structure}, we will
describe the structural properties of the pyrochlores studied.  
Section \ref{bonds} analyses the electronic properties of the 
different bonds, in terms of static and dynamical charges, 
charge density distribution, and decomposition of the partial 
density of states (PDOS).  In Sec.~\ref{discussion}, a discussion 
of our results is presented, and experimental evidences for radiation 
resistance are examined. 
Finally, conclusions are given in Sec.~\ref{conclusions}.

\section{Structural Properties}
\label{structure}

The ideal pyrochlore structure is isometric with space group Fd3m, 
and eight molecules in the unit cell. 
It can be visualized in a variety of ways\cite{review}, 
but is often considered as an ordered defect fluorite structure.  
The A and B metal cations occupy the 16d 
($\frac{1}{2},\frac{1}{2},\frac{1}{2}$) and 16c ($0,0,0$) sites respectively, 
and the oxygens are in the 48f ($x,\frac{1}{8},\frac{1}{8}$) and 8b 
($\frac{3}{8},\frac{3}{8},\frac{3}{8}$) positions.  The anion sublattice
could be completed adding the missing oxygen in the 8a site to form the
fluorite structure.
The system is completely described in terms of the lattice size $a$,
and $x$ parameter defining the position of the O$_{\text{48f}}$.
Geometrically, the smaller B$^{+4}$ cation is surrounded by six O$_{48f}$ 
in a distorted octahedron, and the large A$^{+3}$ cation is in a 
distorted cubic polyhedron formed by two O$_{8b}$ and six O$_{\text{48f}}$.
The 48f oxygen is coordinated to two A and two B sites, and the
8b oxygen is inside a tetrahedron formed by A$^{+3}$ cations. 

We have used a variable cell conjugate-gradient minimization of the energy
following the forces and stresses to obtain a relaxed minimum-energy 
structure for each of the studied systems.
First principles calculations are performed with the self-consistent 
{\sc siesta} method,~\cite{siesta} using Density Functional Theory 
(DFT)\cite{DFT} within the Local Density Approximation (LDA).~\cite{ca}
Norm-conserving pseudopotentials\cite{tm2} in the Kleinman-Bylander 
form\cite{kb} are generated with the following atomic valence 
configurations: Ti(3s$^2$,3p$^6$,3d$^2$,4s$^2$), 
Y(5s$^2$,4p$^6$,4d$^1$), Zr(4s$^2$,4p$^6$,4d$^2$,5s$^2$), 
Sn(5s$^2$,5p$^2$), La(5s$^2$,5p$^6$,5d$^1$,6s$^2$), and Hf(6s$^1$,5d$^3$).  
These, include nonlinear partial-core corrections\cite{PCC} and/or 
semicore states to account for the large overlap between core and 
valence states.  We used a single-$\zeta$ basis set for the semicore states
and double-$\zeta$ plus polarization for the valence states.
The charge density is projected on a real space uniform grid with an
equivalent plane-wave cutoff of 280 Ry, to calculate the
exchange-correlation and Hartree matrix elements in the Hamiltonian.  
Brillouin zone summations are carried out with a 3$\times$3$\times$3 
Monkhorst-Pack k-point mesh.

\subsection{Structural parameters}

The relaxed lattice sizes, and $x$ parameters are given in Table \ref{lattice}.
In general, the calculated LDA lattice parameters are underestimated, 
but the agreement with available experimental 
data\cite{Tabira2000,Kennedy97,Lian03,Lumpkin04} is better than $\sim 1\%$.
There is also good agreement with other first principles 
calculations.~\cite{Chartier02,Panero04}  However small deviations exist
when $x$ is compared to the values calculated using classical models. 
These calculations predict an increase in $x$ with the B cation 
radius\cite{Minervini02}.  
We do not observe this clear trend, although the differences appear mainly 
for the stannates, where the covalent character of tin makes the 
definition of ionic radii, at least, questionable.  To further validate 
the calculated structures for these pyrochlores, we show in 
Tab.\ref{lattice} the most relevant bond distances and angles.
The agreement again is remarkable with deviations to experimental 
values not bigger than $\sim 2\%$.

\begin{table}[b]
\caption{Calculated lattice parameters for A$_2$B$_2$O$_7$ pyrochlores,
compared to experimental and calculated values. Th.A shows results for other
first principles simulations.  Th.B shows results obtained with 
empirical potentials.}
\begin{ruledtabular}
\begin{tabular*}{10cm}{l@{\extracolsep{\fill}}ccccccc}
                  &&   a   &     x     & $d_{A-O_{8a}}$ & $d_{A-O_{\text{48f}}}$ & $d_{B-O_{\text{48f}}}$ & $\widehat{BOB}$\\
\hline
\multicolumn{8}{c}{La$_2$Sn$_2$O$_7$}\\
{\it\small This work}        && 10.744 & 0.3341& 2.33 & 2.61 & 2.10 & 129.1   \\
Exp.[\onlinecite{Kennedy97}] && 10.703 & 0.329 & 2.32 & 2.63 & 2.07 & 131.6 \\
Th.B[\onlinecite{Tabira2000}]    &&        & 0.324 \\
\multicolumn{8}{c}{La$_2$Zr$_2$O$_7$} \\
{\it\small This work}         &&10.724 & 0.3314& 2.32 & 2.62 & 2.09 & 130.5   \\
Exp.[\onlinecite{Lumpkin04}] &&10.737 & 0.331 & \\
Exp.[\onlinecite{Tabira2000}] &&10.805 & 0.333 & 2.34 & 2.64 & 2.10 & 130.6 \\
Th.A[\onlinecite{Chartier02}] && 10.986 & 0.330 &    \\
Th.B[\onlinecite{Tabira2000}]     &&        & 0.326 \\
\multicolumn{8}{c}{La$_2$Hf$_2$O$_7$}\\
{\it\small This work}        && 10.673 & 0.3298& 2.31 & 2.62 & 2.07 & 131.4   \\
Exp.[\onlinecite{Lumpkin04}] &&10.750 & 0.331 & \\
\multicolumn{8}{c}{La$_2$Ti$_2$O$_7$} \\
{\it\small This work}        && 10.366 & 0.3227& 2.24 & 2.59 & 1.98 & 135.6   \\
Th.A[\onlinecite{Chartier02}] && 10.541 & 0.323 & 2.28 & 2.63 & 2.02 & 135.1 \\
\multicolumn{8}{c}{Y$_2$Sn$_2$O$_7$} \\
{\it\small This work}       && 10.348 & 0.3437& 2.24 & 2.44 & 2.07 & 124.1   \\
Exp.[\onlinecite{Kennedy97}]&& 10.372 & 0.337 & 2.25 & 2.49 & 2.04 & 127.6 \\
Th.A[\onlinecite{Panero04}] && 10.329 & 0.338 &    \\
Th.B[\onlinecite{Tabira2000}]   &&        & 0.329 \\
\multicolumn{8}{c}{Y$_2$Zr$_2$O$_7$}\\
{\it\small This work}       && 10.335 & 0.3417& 2.24 & 2.45 & 2.06 & 125.1   \\
Th.A[\onlinecite{Panero04}]  && 10.463 & 0.342 & 2.26 & 2.48 & 2.08 & 125.0 \\
\multicolumn{8}{c}{Y$_2$Hf$_2$O$_7$} \\
{\it\small This work}       && 10.300 & 0.3399& 2.23 & 2.46 & 2.04 & 126.1   \\
\multicolumn{8}{c}{Y$_2$Ti$_2$O$_7$} \\
{\it\small This work}       &&  9.974 & 0.3321& 2.16 & 2.43 & 1.94 & 130.2 \\
Exp.[\onlinecite{Lian03}]   && 10.100 & 0.330 & 2.18 & 2.46 & 1.96 & 131.3 \\
Th.A[\onlinecite{Panero04}] && 10.049 & 0.329 &    \\
Th.B[\onlinecite{Tabira2000}]   &&        & 0.325 \\
\end{tabular*}
\end{ruledtabular}
\label{lattice}
\end{table}

\subsection{Elastic constants}

The linear elastic constants can be obtained from first principles
with the derivatives of the stress as a function of a properly chosen
lattice distortion $\delta$ parametrizing the strain.~\cite{RMartin83}  
For a cubic structure, the number of independent elastic constants is 
reduced to three: $c_{11}$, $c_{12}$ and $c_{44}$.  They can be obtained
through volume compression, tetragonal and trigonal strains:~\cite{Stadler96}

\begin{eqnarray}
\nonumber
\epsilon_{cmp} = \frac{1}{3} \left( \begin{array}{ccc}
\delta & 0 & 0 \\
 0 & \delta & 0 \\
 0 & 0 & \delta \\
 \end{array} \right) 
 &\longrightarrow & \frac{\partial\sigma_{11}}{\partial\delta}=B=\frac{c_{11}+2c_{12}}{3}
 \\
\nonumber
\epsilon_{tet} = \frac{1}{2} \left( \begin{array}{ccc}
-\delta & 0 & 0 \\
 0 & -\delta & 0 \\
 0 & 0 & 2\delta \\
 \end{array} \right) 
 &\longrightarrow & \frac{\partial\sigma_{11}}{\partial\delta}=C'=\frac{1}{2}(c_{11}-c_{12})
 \\
\nonumber
\epsilon_{tri} = \left( \begin{array}{ccc}
\delta^2 & \delta & \delta \\
 \delta & \delta^2 & \delta \\
 \delta & \delta & \delta^2 \\
 \end{array} \right) 
 &\longrightarrow & \frac{\partial\sigma_{21}}{\partial\delta}=2c_{44}
\end{eqnarray}
where $B$ and $C'$ are the bulk modulus and the shear modulus.  
When the applied strain reduces the symmetry of the crystal structure
adding new internal degrees of freedom for the atomic positions, these
have to be fully relaxed to obtain the elastic constants.
In Table \ref{elastic} we show the results for the elastic constants 
and bulk modulus, both for the distorted cells without geometry relaxations 
(c$_{ij}^0$), and for those structures resulting in internal atomic relaxations.  
It can be seen that the relaxation has a strong effect on the elastic
constants, specially for $c_{44}$ that is softened up to a 40\% when the
atomic positions are optimized.  To the best of our knowledge, 
no experimental data is available for the pyrochlores studied in this work.  
It has to be considered that the values presented in Table \ref{elastic} 
were obtained at $T=0$K, and that temperature effects generally reduce 
the values for the elastic constants.  Consequently, we expect the 
experimental values at room temperature to be smaller than the values 
presented here.

\begin{table}[t]
\caption{Calculated elastic constants (in GPa) for A$_2$B$_2$O$_7$ pyrochlores.}
\begin{ruledtabular}
\begin{tabular*}{9cm}{c@{\extracolsep{\fill}}cccccccc}
       & c$_{11}^0$ & c$_{12}^0$ & c$_{44}^0$ & B$^0$ 
       & c$_{11}$ & c$_{12}$ & c$_{44}$ & B \\
\hline
 La$_2$Ti$_2$O$_7$ & 336 & 148 & 309 & 211 & 241 & 196 & 213 & 211 \\
 Y$_2$Ti$_2$O$_7$  & 388 & 160 & 318 & 236 & 381 & 153 & 241 & 229 \\
 La$_2$Zr$_2$O$_7$ & 313 & 164 & 305 & 213 & 290 & 156 & 200 & 200 \\
 Y$_2$Zr$_2$O$_7$  & 394 & 170 & 312 & 245 & 351 & 163 & 198 & 225 \\
 La$_2$Hf$_2$O$_7$ & 339 & 166 & 327 & 224 & 322 & 149 & 214 & 207 \\
 Y$_2$Hf$_2$O$_7$  & 403 & 191 & 343 & 262 & 379 & 167 & 221 & 238 \\
 La$_2$Sn$_2$O$_7$ & 295 & 131 & 280 & 185 & 295 & 131 & 183 & 185 \\
 Y$_2$Sn$_2$O$_7$  & 354 & 141 & 289 & 212 & 348 & 140 & 202 & 209 \\
\end{tabular*}
\end{ruledtabular}
\label{elastic}
\end{table}

\begin{figure}[b]
\includegraphics[scale=0.43]{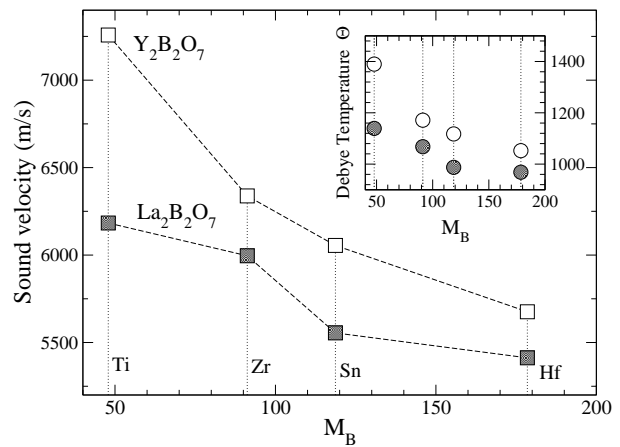}
\caption{Calculated mean sound velocity as a function of the atomic mass 
of the B site for the two families of pyrochlores studied: 
Y$_2$B$_2$O$_7$ (empty symbols) and La$_2$B$_2$O$_7$ (filled symbols).
The Debye temperatures  $\Theta$ (in K) are shown in the inset.}
\label{debye}
\end{figure}

From the elastic constants we can obtain some information on the thermal
properties of the pyrochlores.  The longitudinal and transverse acoustic
velocities (along the [100] direction) are estimated with 
$v_{l}=\sqrt{c_{11}/\rho}$ and $v_{t}=\sqrt{c_{44}/\rho}$, 
where $\rho$ is the mass density of the material.~\cite{Ibach} 
We can then define the mean sound velocity as:
\begin{eqnarray}
\nonumber
\frac{1}{v^3}=\frac{1}{3}\left(\frac{1}{v_l^3}+\frac{2}{v_t^3}\right)
\end{eqnarray}
We observe that the resulting velocities (shown in Fig.\ref{debye}) 
decrease with the atomic mass of the B cation, and also with the mass of A.
We do not observe a clear trend with the cationic radii.
These velocities can also be used to calculate Debye's characteristic 
temperature\cite{Robie66} $\Theta_D=(hv/k_B)(3N_o/4\pi V)^{1/3}$, 
for a solid with $N_o$ atoms in the volume $V$, and where $h$ is 
Plank's constant, $k_B$ is Boltzmann constant (see inset in Fig.\ref{debye}).

\section{Electronic Properties}
\label{bonds}

The concept of {\it atomic charge} is widely used in chemistry 
as well as in solid state physics.  Unfortunately, it is not a well
defined concept. Many different
definitions have been proposed and there is not a general 
agreement in how to partition the electronic charge density around
the atoms and quantify the atomic charge.  
Behind this technical problem lies a more fundamental one:
the quantification of the covalency/ionicity of chemical bonds.
Here we present the results obtained with several definitions
of atomic charges, in order to characterize the nature of the 
different bonds in the family of pyrochlores.

\begin{table*}[t!]
\caption{Calculated dynamical effective charge tensors for the
A ion centered at ($\frac{1}{2},\frac{1}{2},\frac{1}{2}$), B ion
at position ($0,0,0$), O$_{48f}$ ion at ($x,\frac{1}{8},\frac{1}{8}$), and
O$_{8b}$ ion at ($\frac{3}{8},\frac{3}{8},\frac{3}{8}$). 
The effective charge tensor for any other atom can be obtained
applying the symmetry operations of the crystal.  The eigenvalues of the
symmetric part of the tensor are also given.}
\begin{ruledtabular}
\begin{tabular}{ccccc}
\\
& \multicolumn{1}{c}{Z$^*_{A}$}        &    
  \multicolumn{1}{c}{Z$^*_{B}$}         & 
  \multicolumn{1}{c}{Z$^*_{O_{\text{48f}}}$}   &  Z$^*_{O_{8b}}$ \\
La$_2$Sn$_2$O$_7$ &
$\left(\begin{matrix}
{  4.12  &  0.19  &  0.19  \\
   0.19  &  4.12  &  0.19  \\
   0.19  &  0.19  &  4.12 } \end{matrix}\right)$   
& $\left(\begin{matrix}
                     {  4.19  & -0.19  & -0.19  \\
                       -0.19  &  4.19  & -0.19  \\
                       -0.19  & -0.19  &  4.19 } \end{matrix}\right)$ 
&  $\left(\begin{matrix}
                     { -2.24 & 0.00  & 0.00  \\
                        0.00 & -2.31 &  0.24 \\
                        0.00 &  0.24 & -2.31 } \end{matrix}\right)$ 
& -2.88 \\
 & [3.93, 3.93, 4.50] & [4.38, 4.38, 3.81] & [-2.07, -2.24, -2.55] & \\
\\
La$_2$Hf$_2$O$_7$ &
$\left(\begin{matrix}
{  4.09  &  0.23  &  0.23  \\
   0.23  &  4.09  &  0.23  \\
   0.23  &  0.23  &  4.09  } \end{matrix}\right)$ 
&  $\left(\begin{matrix}
                     {  5.69  & -0.40  & -0.40  \\
                       -0.40  &  5.69  & -0.40  \\
                       -0.40  & -0.40  &  5.69  } \end{matrix}\right)$ 
&  $\left(\begin{matrix}
                     { -2.55 & 0.00  & 0.00  \\
                        0.00 & -2.90 &  0.76 \\
                        0.00 &  0.76 & -2.90 } \end{matrix}\right)$ 
&-2.88 \\
 & [3.86, 3.86, 4.55] & [6.09, 6.09, 4.89] & [-2.14, -2.55, -3.66] & \\
\\
La$_2$Zr$_2$O$_7$ &
$\left(\begin{matrix}
{  4.11  &  0.20  &  0.20  \\
   0.20  &  4.11  &  0.20  \\
   0.20  &  0.20  &  4.11  } \end{matrix}\right)$ 
&  $\left(\begin{matrix}
                     {  6.00  & -0.37  & -0.37  \\
                       -0.37  &  6.00  & -0.37  \\
                       -0.37  & -0.37  &  6.00  } \end{matrix}\right)$ 
&  $\left(\begin{matrix}
                     { -2.55 & 0.00  & 0.00  \\
                        0.00 & -3.07 &  1.06 \\
                        0.00 &  1.06 & -3.07 } \end{matrix}\right)$ 
&-2.89 \\
 & [3.91, 3.91, 4.55] & [6.46, 6.46, 5.35] & [-2.01, -2.55, -4.13] & \\
\\
La$_2$Ti$_2$O$_7$ &
$\left(\begin{matrix}
{  4.20  &  0.20  &  0.20  \\
   0.20  &  4.20  &  0.20  \\
   0.20  &  0.20  &  4.20  } \end{matrix}\right)$ 
&  $\left(\begin{matrix}
                     {  6.91  &  0.01  &  0.01  \\
                        0.02  &  6.90  &  0.01  \\
                        0.02  &  0.01  &  6.90  } \end{matrix}\right)$ 
&  $\left(\begin{matrix}
                     { -2.57 & 0.00  & 0.00  \\
                        0.00 & -3.53 &  1.56 \\
                        0.00 &  1.56 & -3.53 } \end{matrix}\right)$ 
&-2.99 \\
 & [4.00, 4.00, 4.60] & [6.89, 6.89, 6.93] & [-1.97, -2.57, -5.09] & \\
\\
Y$_2$Sn$_2$O$_7$ &
$\left(\begin{matrix}
{  3.93  &  0.06  &  0.06  \\
   0.06  &  3.93  &  0.06  \\
   0.06  &  0.06  &  3.91  } \end{matrix}\right)$ 
&  $\left(\begin{matrix}
                     {  4.06  & -0.18  & -0.18  \\
                       -0.18  &  4.07  & -0.18  \\
                       -0.18  & -0.18  &  4.06  } \end{matrix}\right)$ 
&  $\left(\begin{matrix}
                     { -2.15 &  0.00 &  0.00 \\
                        0.00 & -2.24 &  0.18 \\
                        0.00 &  0.18 & -2.24 } \end{matrix}\right)$ 
&-2.69 \\
 & [3.86, 3.87, 4.04] & [4.24, 4.24, 3.70] & [-2.06, -2.15, -2.42] & \\
\\
Y$_2$Hf$_2$O$_7$ &
$\left(\begin{matrix}
{  3.90  &  0.07  &  0.07  \\
   0.07  &  3.90  &  0.07  \\
   0.07  &  0.07  &  3.90  } \end{matrix}\right)$ 
&  $\left(\begin{matrix}
                     {  5.67  & -0.45  & -0.45  \\
                       -0.45  &  5.67  & -0.45  \\
                       -0.45  & -0.45  &  5.67  } \end{matrix}\right)$ 
&  $\left(\begin{matrix}
                     { -2.55 &  0.00 &  0.00 \\
                        0.00 & -2.84 &  0.74 \\
                        0.00 &  0.74 & -2.84 } \end{matrix}\right)$ 
&-2.68 \\
 & [3.83, 3.83, 4.04] & [6.12, 6.12, 4.77] & [-2.10, -2.55, -3.58] & \\
\\
Y$_2$Zr$_2$O$_7$ &
$\left(\begin{matrix}
{  3.94  &  0.04  &  0.04  \\
   0.04  &  3.94  &  0.04  \\
   0.04  &  0.04  &  3.94  } \end{matrix}\right)$ 
&  $\left(\begin{matrix}
                     {  5.98  & -0.46  & -0.43  \\
                       -0.46  &  5.98  & -0.43  \\
                       -0.43  & -0.43  &  5.98  } \end{matrix}\right)$ 
&  $\left(\begin{matrix}
                     { -2.55 &  0.00 &  0.00 \\
                        0.00 & -3.01 &  1.03 \\
                        0.00 &  1.03 & -3.01 } \end{matrix}\right)$ 
&-2.70 \\
 & [3.90, 3.90, 4.02] & [6.40, 6.44, 5.10] & [-2.00, -2.55, -4.02] & \\
\\
Y$_2$Ti$_2$O$_7$ &
$\left(\begin{matrix}
{  3.96  &  0.03  &  0.03  \\
   0.03  &  3.96  &  0.03  \\
   0.03  &  0.03  &  3.96  } \end{matrix}\right)$ 
&  $\left(\begin{matrix}
                     {  6.89  & -0.16  & -0.16  \\
                       -0.16  &  6.89  & -0.16  \\
                       -0.16  & -0.16  &  6.89  } \end{matrix}\right)$ 
&  $\left(\begin{matrix}
                     { -2.57 &  0.00 &  0.00 \\
                        0.01 & -3.46 &  1.50 \\
                        0.01 &  1.50 & -3.46 } \end{matrix}\right)$ 
&-2.73 \\
 & [3.93, 3.93, 4.02] & [7.05, 7.05, 6.57] & [-1.96, -2.57, -4.96] & \\
\end{tabular}
\end{ruledtabular}
\label{BC}
\end{table*}

\subsection{Dynamical charges}
The so called {\it dynamical charges} (or Born charges) are defined in 
terms of the change of the polarization created when an atom is 
displaced from its equilibrium position.  They have been used to
characterize the ionicity/covalency in ferroelectric 
perovskites.~\cite{Posternak94} The tensorial definition is:
\begin{equation}
Z_{i,\alpha\beta}^*=\Omega_0\frac{\partial P_\beta}{\partial u_{i,\alpha}}
\end{equation}
where $P_\beta$ is the $\beta$ component of the macroscopic polarization 
induced per unit cell when the atom $i$ is displaced in the direction 
$u_\alpha$, and $\Omega_0$ is the unit cell volume.

The Born effective charges for the four inequivalent lattice points are
shown in Table \ref{BC}.  The charge neutrality sum rule ($\sum Z^*=0$) 
is frequently used to benchmark the accuracy of the calculation, 
and in our case is fulfilled to within 0.04$e$.  
The symmetry of the A and B sites allow for a decomposition of the 
Born charge tensors into two independent eigenvalues, one for 
displacements pointing in the direction $[111]$ and the others (degenerate) for 
displacements in the orthogonal plane.  For the A site, the highest eigencharge
corresponds to the $[111]$ direction, due to the presence of the O$_{8b}$ along 
this axis.  On the other side, the $[111]$ gives the smallest eigencharge 
because the O$_{8a}$ site is unoccupied in pyrochlores.
Small deviations from the two-fold degenerate eigencharges are observed 
for some systems, due to the unconstrained geometry relaxation.
For the O$_{48f}$ position, the diagonalization of the symmetric
part of the Born tensor gives three non-degenerate eigenvectors,
along the $[011]$, $[100]$ and $[0\bar{1}1]$ directions 
(with increasing eigenvalues in the same order).  If we consider the 
plane defined by the B--O--B elements, the first direction corresponds to the
C$_2$ axis, the second is perpendicular to the plane, and the third goes from 
B to B.  Finally, the O$_{8b}$ tensor is isotropic, and only the diagonal 
component is given in the table.

The charge for the A site (generally higher than the nominal +3$e$, 
both for the La and Y pyrochlores studied here) is almost independent 
of the B site composition, with variations smaller than 6\%.  
Similar variations ($\sim$10\%) are observed for the O$_{8b}$.  
A larger chemical dependence is observed for the dynamical charges
associated to the octahedron formed by the B cation and the O$_{\text{48f}}$.
In this case, there are differences in the Born charges for the 
B cation of up to 40\%, and up to 30\% for the O$_{\text{48f}}$, 
when the B-site is occupied by Sn instead of Zr or Ti.

The dynamical charges $Z_B^*$ increase in the order B=Sn, Hf, Zr, Ti.  
The large anomalous charge contributions (additional charges to the 
nominal ionic values of A$^{+3}$, B$^{+4}$ and O$^{-2}$) originate from
the electronic charge reorganization induced when an atom is displaced 
from its original position.  They are correlated to charge transfers 
between atoms and dynamical changes in the hybridization.
Large values of Born charges have been observed in Ti-perovskites and
attributed to the hybridization between the occupied O $2p$ orbitals and 
the unoccupied Ti $3d$ orbitals.~\cite{Ghosez98}. Hence, it would be naive
to conclude that the degree of ionicity in pyrochlores increases in 
the same direction (Sn, Hf, Zr, and Ti) as $Z_B^*$.

\begin{figure}[b]
\includegraphics[scale=0.50]{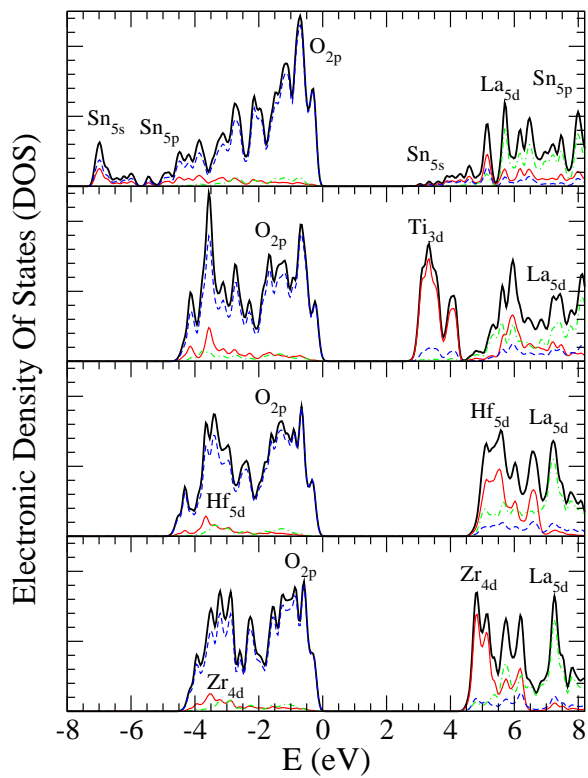}
\caption{({\it Color online}) Electronic density of states (DOS) 
for the La$_2$B$_2$O$_7$ pyrochlores, with B=Sn, Ti, Hf and Zr 
(from top to bottom panel).  The top of the valence band is at 0 eV. The 
dashed, and dot-dashed lines show the projected contribution 
into the O$_{2p}$ and La$_{5d}$ atomic orbitals.  The thin solid line
shows the projection into the Sn $5s$ and $5p$ orbitals, the
Ti $3d$, Zr $4d$, and Hf $5d$, for each system.}
\label{DOS}
\end{figure}

Figure \ref{DOS} shows the density of states close to the Fermi level 
for the four lanthanum structures studied (those for yttrium-pyrochlore 
are very similar). The valence band has a main contribution from the 
$2p$ orbitals of oxygen, and a small contribution from the $d$ orbitals 
of the B cation ($5s$ and $5p$ orbitals for Sn).  The conduction bands
have a sharp peak with B cation $d$ orbital character in its lower part, 
and contributions from the $5d$ orbitals of the La atom.  
La$_2$Sn$_2$O$_7$ has a very dispersive conduction band with 
tin-$5s$ character.  The anomalous contributions for the B cations can be
interpreted as a result of electron current flowing from the occupied 
O$_{2p}$ states into the empty states of the B-cation in the conduction band
when the B ion is displaced from its equilibrium position.
The calculated band gaps are 2.6, 2.8, 4.7 and 4.5 eV for Sn, Ti, Hf, 
and Zr respectively.  The LDA band gaps are known to be underestimated, 
and a correction of the strong correlation effects is likely to increase 
the band gap and reduce the Born effective charges\cite{Filippetti03}.  

\subsection{Static charges}
Many definitions of atomic static charges can be found in the literature. 
They differ in the way the space is partitioned and the charge is 
associated to each atom.  Probably the most popular method is the 
Mulliken population analysis\cite{Mulliken}, where the atomic basis 
functions are used to associate the electronic density around each atom.
Unfortunately, it is well known that the atomic charges obtained with this
method are basis-dependent.  Recently, Fonseca-Guerra {\it et. al.} 
showed that the Voronoi Deformation Density (VDD) method give chemically
meaningful charges with basis set independency\cite{Voronoi}.  
This method is based in direct integration of the deformation 
density $\delta\rho(r)=\rho(r)-\rho_{atom}(r)$ over the Voronoi polyhedron 
around each atom. The charges obtained with VDD are similar to the values 
obtained with the Hirshfeld method\cite{Hirshfeld}, where the electron 
density is assigned to each atom by weighting with the isolated atom 
density $\rho_{atom}(r)$.

\begin{figure}
\includegraphics[scale=0.53]{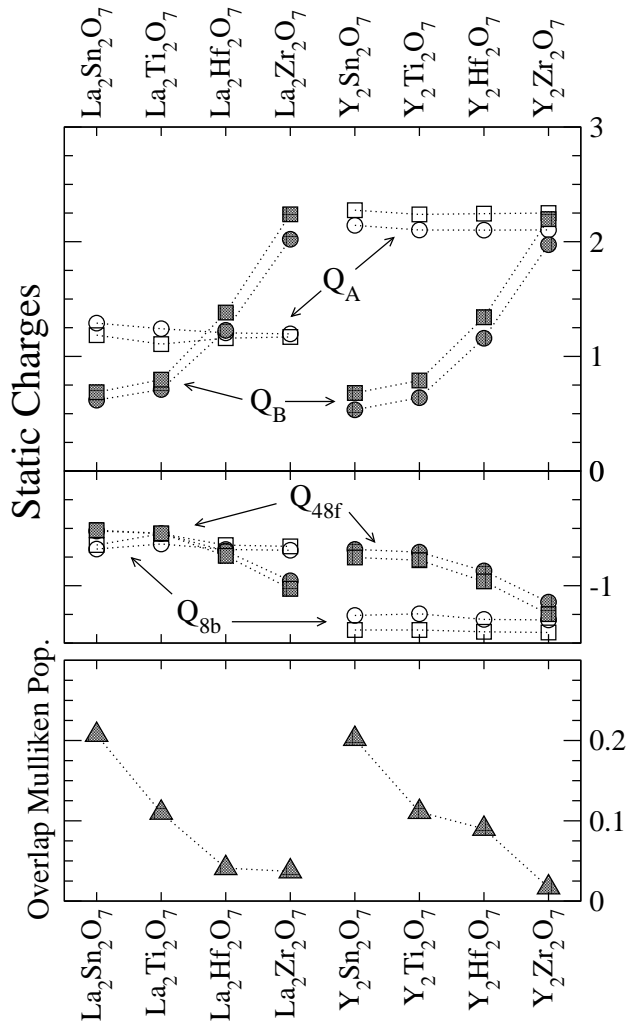}
\caption{Static charges calculated with VDD (squares) and Hirshfeld 
(circles) methods.  Upper panel: the cationic charges for the 
A and B sites (empty and filled symbols). 
Middle panel: oxygen 48f (filled) and 8b charges (empty).
Lower panel: Overlap Mulliken populations between B and O$_{\text 48f}$
for the two families studied.}
\label{static}
\end{figure}

In Fig.\ref{static} we show the calculated static charges with VDD and
Hirshfeld methods for the cations A and B, and for the two anion sites 
8b and 48f.  We observe that the values of Hirshfeld and Voronoi charges 
are very similar.  The composition of the A site affects slightly the 
values for other atoms, specially for the O$_{8b}$.  This is because the 
8b oxygens are surrounded by just four A cations, and no B cation is close 
enough.  The most important dependence on the chemical composition appears in 
the charge of the cation B site, with increasing values in the order 
Sn, Ti, Hf and Zr.  If the degree of ionicity of the pyrochlores 
is to be associated to the static atomic charges this very ordering
would be the order of increasing ionicity (larger atomic charge).
Notice that the O$_{\text 48f}$ anions also accumulate more 
negative charge as we move from Sn to Zr, showing that the B-O$_{\text 48f}$ 
pairs become more ionic.  To further validate this assignment, we also 
show the {\it overlap Mulliken population} between the oxygen 
atoms and the atoms in the A and B sites~\cite{Lumpkin05} (lower panel).
These values, though basis-dependent, can be used to estimate the 
degree of covalency (the larger the overlap, the larger the covalency).  
It is shown that the overlap population of the oxygen decreases in 
the order Sn$>$Ti$>$Hf$>$Zr. 

\begin{figure*}
\includegraphics[scale=0.80]{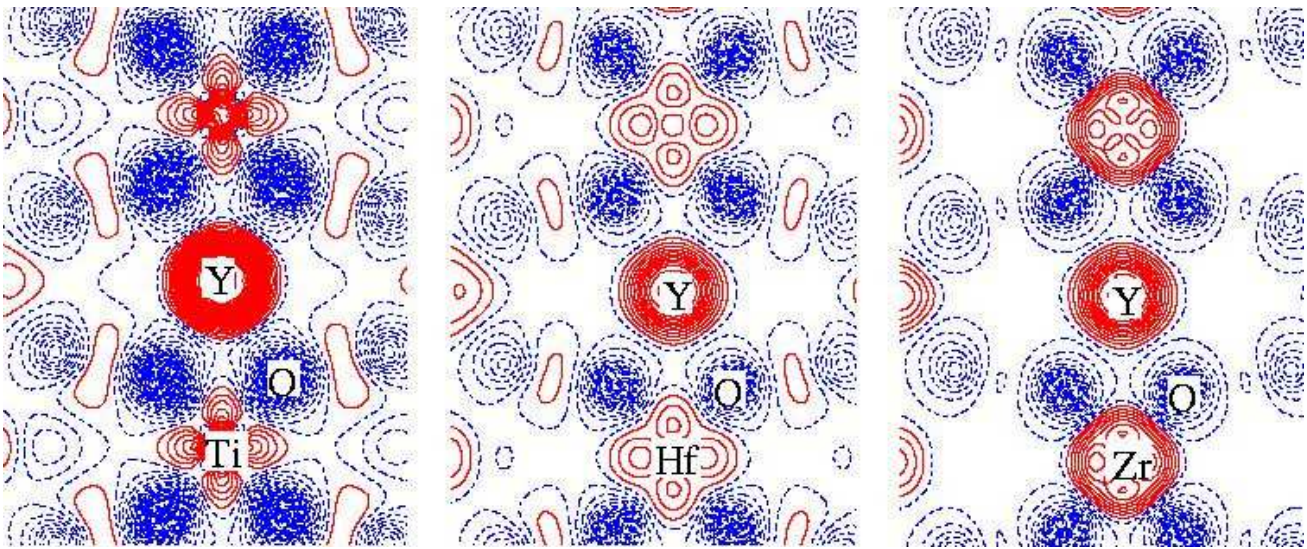}
\caption{(Color online) Electronic charge redistribution in the 
A-B-O$_{48f}$ plane.  The lines are taken at intervals of 0.005$e$ 
in the range -0.25$e$ to 0.25$e$. Solid (red) and dashed (blue) lines
correspond to positive and negative values respectively.}
\label{contour}
\end{figure*}

\subsection{Charge distribution}

Figure \ref{contour} plots the contour of electronic charge redistribution 
$\delta\rho(r)$ in a plane defined by A-B-O$_{48f}$, for Y$_2$Ti$_2$O$_7$,
Y$_2$Hf$_2$O$_7$, and Y$_2$Zr$_2$O$_7$.  The valence configuration for 
Ti, Hf and Zr is similar, as well as the lattice structure.  Consequently, 
the differences in the contour plots have to be attributed to the difference
bonding character with oxygen in the three systems.  The stannate is not 
included here, because of the different valence configuration of Sn.
The directional character of the Ti--O bond is in contrast with the mainly 
ionic Zr--O.  The hafnate presents an intermediate directionality.  
Similar contour plots are observed for the lanthanides.  
Notice that the charge distribution around the A site is mainly 
spherical showing the high ionicity.  The contour plots support the
hypothesis of increasing covalency from Ti to Hf to Zr based pyrochlores.

\section{Discussion}
\label{discussion}

Understanding response of materials to radiation is a very complex problem.
Many factors have been named to be relevant in the context of resistance 
to amorphization by irradiation\cite{Trachenko}, including  topological 
freedom, glass-forming ability, melting and crystallization 
temperatures, ionicity, bond energy, elasticity, ratio of 
ionic radii, defect formation energies, etc.  Many criteria to characterize
radiation resistance focus in a particular subset of properties.
In pyrochlores, the defect-formation energies have been a popular criterion. 
It was thought that the ionic radius or ionic radius ratio were the most 
important factors in determining the defect energies,~\cite{Sickafus00} 
but recent first-principles calculations have demonstrated that the 
electronic structure plays an important role.~\cite{Panero04}
The fact that all pyrochlores have the same structure, makes them an 
excellent playground to tests the relevance of some of these factors, and
particularly the role of the bond-type.  

Naguib and Kelly proposed a criterion for resistance to amorphization by 
irradiation based on the good empirical correlation between ionicity and
susceptibility to amorphization\cite{Naguib75}.  
Frequently, the ionicity of the material
is quantified from the difference in Pauling or Phillips electro-negativities 
for different atoms resulting in inconsistencies of the Naguib and Kelly 
criterion.  Nevertheless, the importance of ionicity/covalency in the
radiation resistance is evident.  For the family of pyrochlores studied 
here, the static atomic charges, and overlap Mulliken populations 
indicate that ionicity increases as the B site is occupied by 
Sn, Ti, Hf and Zr.  Remarkably, this is consistent with the increase 
of resistance of these materials.  The critical temperature (T$_c$) above 
which complete amorphization of the material can not be obtained is frequently 
used as a measure of the resistance to amorphization of the material.  The
lower T$_c$, the more resistant the material.  For the family of lanthanide
pyrochlores La$_2$Sn$_2$O$_7$ (1025 K), La$_2$Hf$_2$O$_7$ (563 K) 
and La$_2$Zr$_2$O$_7$ (339 K), the trend agrees with the decrease of 
covalency, and with the Naguib and Kelly criterion.  Furthermore, the 
observed\cite{Wang99} increase in resistance to amorphization with 
increasing zirconium content in Gd$_2$Ti$_{2-x}$Zr$_x$O$_7$ seems to 
follow the same trend.
However, Y$_2$Sn$_2$O$_7$ with a critical temperature similar to the one 
observed for the hafnate, deviates from this trend and would require 
further study.

The deformation density contour plot for A$_2$Ti$_2$O$_7$ indicates 
that these pyrochlores have an important 
covalent character, as opposed to A$_2$Hf$_2$O$_7$ and A$_2$Zr$_2$O$_7$
where a more ionic character can be deduced from the deformation density.
The nearly identical defect formation energies calculated for the Ti and Sn
members in reference [\onlinecite{Panero04}] could be a consequence of 
the similar covalent bond.  We believe that the highly anomalous Born 
charge calculated for Ti has an origin on the hybridization 
between the $2p$ orbitals of oxygen and the $3d$ orbitals of Ti, similarly
to the anomalous Born charges observed for example in Ti-perovskites, or 
TiO$_2$, and should not be attributed to a strongly ionic character of 
these pyrochlores.  

The calculated elastic constants do not satisfy the Cauchy relation 
$c_{12}/c_{44}=1$.  This condition has to be satisfied in crystals with
two-body interactions, as in a purely ionic system.  In particular, 
for La$_2$Sn$_2$O$_7$, La$_2$Hf$_2$O$_7$,  Y$_2$Ti$_2$O$_7$ and 
Y$_2$Sn$_2$O$_7$ we have $c_{12}/c_{44}<0.75$. The presence of non-central
forces, such as those coming from covalent bonding produces deviations from
Cauchy equality.  The relation is more strongly broken for the unrelaxed 
elastic constants ($c^0_{12}/c^0_{44}$), and is partially recovered under 
relaxation.  The surprising recovery observed for La$_2$Ti$_2$O$_7$ 
could be related to the instability of the pyrochlore 
structure for this system.

\section{Conclusions}
\label{conclusions}
In this work we have used density functional theory calculations to analyze 
the structural and chemical properties of a family of pyrochlores.  
The lattice and structural parameters ($a$ and $x$) were obtained 
with a conjugate gradient minimization of the forces and strains.  
Applying a particular set of lattice distortions, the elastic constants 
were also obtained.  The structural 
properties obtained from first principles can be used to parametrize 
new interatomic potentials that could be used in classical Molecular Dynamics 
simulations for materials design.  The lack of an accurate description 
of the chemical bond in these classical simulations was criticized 
recently in the context of resistance to amorphization by irradiation.  

Although experimental evidence shows that the electronic structure has 
to play some role in radiation resistance, it is still not clear what is 
the trend.  The similar lattice structure for the ideal systems 
(not all of them stable in the pyrochlore structure) studied here, makes 
them excellent systems to elucidate the relevance of the 
chemical properties.  If the description of Trachenko and collaborators 
is correct, the competition between short (covalent) and long (ionic) 
interactions is strongly related to the resistance to 
amorphization\cite{Trachenko,Trachenko05}.  

In this work we present a study of some indicators of covalency/ionicity in
pyrochlores.  Although some of these parameters (static charges, deformation 
density) seem to show evidence that the stannates and titanates pyrochlores 
have a more covalent character than the highly ionic hafnates and zirconates, 
(which are known to be more resistant to amorphization), others, like the 
dynamical charges, do not show the same trend.  Further work is required to
understand these differences.  In particular, a systematic experimental study of 
T$_c$ is desirable, as most of the available data is recorded in different 
conditions (bombarding ions, energy, etc.).

\acknowledgments
This work was supported by BNFL and UK's NERC through the eminerals 
consortium.  The authors are grateful to G. Lumpkin, K. Whitle, K. Trachenko, 
I. Farnan, and S. R\'{\i}os for fruitful discussions on radiation resistance 
of pyrochlores.  EA thanks the Donostia International Physics Center (DIPC) for
its hospitality.


\begin{thebibliography}{99}
\bibitem{unstable}{Not all the compounds in the families studied are stable in the pyrochlore structure.  For example La$_2$Ti$_2$O$_7$ has the layered perovskite structure.}
\end{thebibliography}
\end{document}